\documentclass[]{pasj01}
\Received{$\langle$reception date$\rangle$}
\Accepted{$\langle$acception date$\rangle$}

\begin{document}
\title{Constraints on the dark mass distribution surrounding Sgr A*:
  simple $\chi^2$ analysis for the redshift of photons from orbiting stars}
\author{
Yohsuke \textsc{Takamori}, \altaffilmark{1}$^{*}$
Shogo \textsc{Nishiyama},\altaffilmark{2}
Takayuki \textsc{Ohgami},\altaffilmark{3}
Hiromi \textsc{Saida},\altaffilmark{4}
Rio \textsc{Saitou},\altaffilmark{4}
Masaaki \textsc{Takahashi},\altaffilmark{5}}
\altaffiltext{1}{National Institute of Technology (KOSEN), Wakayama College, Gobo, Wakayama 
644-0023, Japan}
\altaffiltext{2}{Miyagi University of Education, Sendai, Miyagi 980-0845, Japan}
\altaffiltext{3}{Konan University, Kobe, Hyogo 658-8501, Japan}
\altaffiltext{4}{Daido University, Naogya, Aichi 457-8530, Japan}
\altaffiltext{5}{Aichi University of Education, Kariya, Aichi 448-8542, Japan}

\email{takamori@wakayama-nct.ac.jp}


\maketitle

\begin{abstract}
Sagittarius A* (Sgr A*) is the central supermassive black hole with the mass 
$\sim 4\times 10^6 M_{\odot}$ in the Milky Way and stars are orbiting around it.
In May 2018, one of the nearest stars to Sgr A* named S0-2/S2 experienced the 
pericenter passage.
The redshift of photons from S0-2 had varied from $4000\>{\rm km\>s^{-1}}$ to 
$-2000\>{\rm km\>s^{-1}}$ during the pericenter passage, which is within 
$0.5\>{\rm yr}$. 
In this paper, we show that this steep variation of the redshift gives a strong 
constraint on a dark mass distribution inside the orbit of S0-2. 
By applying a simple $\chi^2$ analysis to the observed redshift, 
we can easily distinguish between
the two models, the point mass model and the point mass plus an 
extended mass model without the best-fitting parameter search.
Our redshift data during the pericenter passage in 2018 with Subaru/IRCS 
bound the amount of the extended 
mass inside the orbit of S0-2 less than $0.5\,\%$ ($\sim 2 \times 10^4 M_\odot$) 
of the mass of Sgr A*. This constraint obtained by our simple analysis 
is comparable to previous works with 
the best-fitting parameter search to the motion of S0-2 including the 
effect of the extended mass. 
We consider both the power-law and the Plummer models for the dark mass 
distribution model, but the significant difference between 
these results is not found. 
\end{abstract}

\section{Introduction}
Sagittarius A* (Sgr A*)  is a radio source which locates at the center of the Milky Way. 
Astronomers have been observing around Sgr A* over the past few decades and found 
that stars move around it. 
The stars, which are called S-stars, tell us that Sgr A* is a  supermassive black hole 
whose mass is about  $4\times 10^6 M_{\odot}$  (see the latest research: 
\cite{Grav18, D19, S19, Grav20}). 
In 2018, we had a big event for Sgr A*. 
One of the nearest S-stars to Sgr A* named S0-2/S2 (hereafter we call it S0-2), experienced 
the pericenter passage (the distance from Sgr A* is about 
$120\>{\rm au}$) in the year.  
Astronomers expect that this big event gives us new information to test Einstein's 
gravity  theory in the environment around a supermassive black hole. 
Three telescopes, which are Keck observatory, Very Large 
Telescope (VLT), and Subaru telescope, have been ready to observe that big 
event in 2018 independently.  Keck and VLT can carry out both astrometric and 
spectroscopic measurements for S-stars  (e.g., \cite{B16}; \cite{G17}). 
Subaru telescope can perform high-resolution spectroscopic measurements with IRCS 
\citep{N18}.  As a result of their astrometric/spectroscopic observations in 2018,
they found that S0-2 experienced the pericenter passage in May 2018. 
Moreover, they showed that the spectroscopic measurements of S0-2 during the 
pericenter passage strongly suggest that Einstein's gravity is preferable to 
Newton's gravity in the environment around a supermassive black hole 
\citep{Grav18, D19}.
In addition to these works, \citet{S19} has pointed out 
another view of the ``general relativistic effect'' within the spectroscopic data
obtained by Subaru/IRCS during the pericenter passage of S0-2 in 2018.
Recently, \citet{Grav20} has performed the orbital fitting to the motion of 
S0-2 with the data up to the end of 2019 and reported that they caught the 
general relativistic pericenter shift in the data.

The present observational data of S0-2 are well explained under the assumption 
that S0-2 feels only the gravity from Sgr A*. 
However, within the uncertainties of the observational data, the deviation from the 
above assumption can be examined. 
As one possibility of such deviation, we consider a dark mass distribution surrounding  
Sgr A*, e.g., faint S-stars, neutron stars, stellar mass black holes, faint accretion gas 
clouds, and a dark matter. 
The amount of such dark extended mass has been bounded within the 
uncertainties of the observational data.
Before the pericenter passage of S0-2 in 2018, 
the acceptable amount of the extended mass inside the orbit of S0-2 is 
1\,\% of the mass of Sgr A* at most \citep{B16, G17}.
After the observations of S0-2 in 2018, 
\citet{Grav18} reported that the acceptable extended 
mass is in the range 0.35\,\%--1\,\% depending on the extended 
mass models, though they did not show their observed data
and the detail of their analysis.
\citet{D19} performed the Beyasian parameter estimation for the motion 
of S0-2 including the profile of the dark mass distribution.
As a result, they obtained the absolute mass of the extended mass within the 
orbit of S0-2, which is  $5.5\times 10^3M_{\odot}/12.7\times 10^3M_{\odot}$ 
with 1$\sigma$/2$\sigma$.
Converting into the percentage with $M_{\rm BH}=4\times10^6 M_{\odot}$,  
it is in the range 0.14\,\%--0.32\,\%.
Moreover, \citet{Grav20} showed that the upper limit of the extended mass 
inside the orbit of S0-2 is $\sim 0.1\>\%$ ($\sim 4\times10^3M_{\odot}$) 
of Sgr A* with $1\sigma$.  

In the previous papers \citep{Grav18, D19, Grav20}, 
they have given the upper limits of the amount of 
the extended mass by performing the best-fit parameter search for the motion
of S0-2 and Sgr A* including a dark mass distribution with 
their astrometric and spectroscopic data of S0-2 during the pericenter passage in 2018.
In this paper, we focus on the spectroscopic data of S0-2 and show 
that the data during the pericenter passage 
give a strong constraint on the amount of the extended mass inside the orbit of S0-2.
The spectroscopic data tell us the redshift of photons from S0-2.
From the observation of S0-2 during the pericenter passage in 2018,
we can see that the redshift had varied from $4000\>{\rm km\>s^{-1}}$ to 
$-2000\>{\rm km\>s^{-1}}$ within $0.5\>{\rm yr}$.
Thanks to this steep variation of the redshift, we can easily distinguish the two models,
the point mass model and the point mass plus an extended  mass model.
The aims of this paper are as follows:
\begin{enumerate}
   \item[(i)] Propose a simple $\chi^2$ analysis which needs much less numerical 
  costs compared with the method used in the previous papers \citep{D19, Grav20}, 
  and give a constraint on a dark mass distribution around Sgr A*.
  \item[(ii)] Show our spectroscopic data obtained by Subaru/IRCS during the pericenter 
 passage of S0-2 in 2018 can give a strong constraint on the amount of 
 the extended mass inside the orbit of S0-2 by using the $\chi^2$ analysis
 proposed in (i).
\end{enumerate}
We do {\it not} perform the best-fitting parameter search including a dark mass 
distribution which needs much numerical cost \citep{Grav18, D19, Grav20}. 
Instead of the best-fitting parameter search, we suggest a simple $\chi^2$ analysis
with spectroscopic data. 
The procedure of our $\chi^2$ analysis is as follows.
We solve the equation of motion of a S-star in two cases of the point mass and 
the point mass plus an extended mass models. 
Then, we calculate the value of $\chi^2$ with spectroscopic data for the 
motion of the S-star in both models and take the ratio of those $\chi^2$.  
Our method relies on the fact that the present observational data are well 
explained by the point mass model, and the effect of the extended mass 
can be regarded as a small perturbation within the uncertainties of 
observed data.  The ratio of $\chi^2$ becomes an indicator whether
the dark mass distribution model is acceptable.
For S0-2, the time evolution of the redshift had varied from 
$4000\>{\rm km\>s^{-1}}$ to $-2000\>{\rm km\>s^{-1}}$ 
during the pericenter passage, which is within $0.5\>{\rm yr}$. 
Thanks to this steep variation of the redshift, 
we can infer how much mass for the dark extended object is acceptable.
By applying our $\chi^2$ analysis to our work \citet{S19},  
we find that the amount of the extended mass inside the orbit of S0-2 is 
less than  $0.5\,\%$ ($\sim 2\times10^{4} M_{\odot}$) of Sgr A*.
Our result is stronger by the factor $1/2$ than the results obtained 
before the pericenter passage \citep{B16, G17}.  
This constraint is comparable to the recent works including 
the data during the pericenter passage in 2018 
\citep{Grav18, D19, Grav20}.
In \citet{Grav18} and \cite{D19}, they considered the power-law model 
\citep{Gh08, B16, G17} for the dark mass distribution which represents 
a stellar cluster and is supported by the surface brightness of the 
galactic center region with infrared observations \citep{Gez03, Sch07}.
In addition to that model, we also adopt a Plummer model 
\citep{R01, M05, G09, Grav20} for the dark mass distribution. 
The Plummer model is applied initially to a globular cluster \citep{P11} 
and is available for a dark mass distribution model around Sgr A*. 
Although we consider these two models, we do not find a significant 
difference between those results.

This paper is organized as follows. In section \ref{sec:2}, we introduce the 
equation of motion of an orbiting star around Sgr A* 
in the post-Newtonian treatment in the context of the general relativity. 
Moreover, we introduce two dark mass distribution models: the power-law model; 
the Plummer model. 
In section \ref{sec:3}, we discuss the influence of the dark mass distribution on the
redshift of photons from S0-2.  The dark mass distribution affects the timing of the 
pericenter passage. It plays an important role to give a strong constraint on the 
dark mass distribution.
In section \ref{sec:4}, we explain our $\chi^2$ analysis to the observed 
redshift of S0-2.
We can easily distinguish the point mass model and the point mass plus 
an extended mass model with our method. 
As a result, we obtain the upper limit of the amount of the extended mass, 
which is $0.5\,\%$ of Sgr A*.
The last section is devoted to the summary and discussion.
Through this paper, $c$ and $G$ represent the speed of light and
Newton's gravity constant, respectively. 

\section{Equation of motion of an orbiting star and dark extended mass models}
\label{sec:2}
\subsection{Equation of motion of an orbiting star in the post-Newtonian approximation}
We calculate the motion of an orbiting star around Sgr A* 
based on general relativity.
Since S-star's orbits are far from Sgr A*(even the periapsis distance of S0-2 is 
about 2000 times the size of the central black hole), we do not need full 
general relativistic treatment.
In the case of S0-2, the post-Newtonian approximation is available.
The equation of motion of a free test particle around a central object 
with mass $M$ can be written as
\begin{equation}
 \frac{d^2{\bf r}}{dt^2}
  =-\frac{GM}{r^3}{\bf r}
  +\frac{GM}{c^2r^3}\left(\frac{4GM}{r}-v^2\right){\bf r}
  +\frac{4GM{\bf r}\cdot{\bf v}}{c^2r^3}{\bf v},
  \label{EOM}
\end{equation}
where ${\bf r}$ is the position vector of the test particle with
respect to the central object, and ${\bf v}$ is its velocity. 
Then, $r$ and $v$ are the absolute value of ${\bf r}$ and 
${\bf v}$, respectively.
The first term on the right-hand side is Newton's gravity and the 
others are the general relativistic effects. 
We need to include influence from a dark mass distribution
surrounding Sgr A* to equation (\ref{EOM}).
The most simple deformation is making the mass $M$ be a function 
of ${\bf r}$, that is $M \rightarrow M({\bf r})$ \citep{R01}.
This deformation would be enough to express a small mass fraction 
surrounding Sgr A*. 
We apply equation (\ref{EOM}) with $M({\bf r})$ to the dynamics of S0-2 
and solve it numerically. 

\subsection{Point mass plus an extended mass model}
Various possibilities are available for a dark mass distribution surrounding Sgr A*.
In this paper, we mainly focus on a stellar cluster model based on 
the surface number density of stars near Sgr A* by infrared observation \citep{Gez03,Sch07}. 
We consider two models representing a spherically symmetric mass distribution;
the power-law model \citep{Gh08, B16, G17, Grav18, D19} and 
the Plummer model (\cite{R01, M05, G09, Grav20}).
We briefly summarize these models here.

Let us show the power-law model at first. The enclosed mass $M(r)$ with 
the mass density proportional to $r^{-\gamma}$ can be written as
\begin{equation}
M(r) = \left\{
\begin{array}{ll} 
  M_{\rm tot}\left\{1-\eta
  +\eta\left(\frac{r}{r_{\rm c}}\right)^{3-\gamma}\right\}, &  
                                                             (r \leq r_{\rm c}); \\
  M_{\rm tot}, & (r>r_{\rm c}),
\end{array} \right. \label{p-law}
\end{equation}
where $M_{\rm tot}$ is the total mass of the central black hole and the dark mass distribution
within $r_{\rm c}$ and $\eta~(0\leq\eta<1)$ is the ratio of the amount of the extended 
mass to the mass of the central black hole. Then, the mass of the central
black hole is given by $M_{\rm BH}=(1-\eta)M_{\rm tot}$. 
The power-law slope $\gamma$ is taken as $2.5$ in \citet{Grav18} or as $1.5$ in \citet{D19}. 
\citet{Gh08} showed that the extended mass upper limit did not depend strongly on the value 
of $\gamma$ in the range from $0.5$ to $3$. 
In this paper, we choose $\gamma$ to be $1.5$ as in \citet{D19}.
Since the cutoff radius is taken as  $r_{\rm c}=0.011\>{\rm pc}$ in \citet{D19}, 
we also use $0.011\>{\rm pc}$ for $r_{\rm c}$.\footnote{
\citet{B16} set the cutoff radius to $r_{\rm c}=0.011\>{\rm pc}$, such that 
it encloses the orbits of S0-2 and S0-38.}

Next, let us show the Plummer model which is applied initially 
to a globular cluster \citep{P11}. 
The mass function of the Plummer model is given by
\begin{equation}
  M(r) 
  = M_{\rm tot}
  \left(1-\eta+\eta
    \frac{\int_0^{r}\tilde{\rho}(\xi)\xi^2d\xi}{\int_0^{r_{0}}\tilde{\rho}(\xi)\xi^2d\xi}
  \right),
\label{plummer}
\end{equation}
where $r_0$ is a parameter which makes $M_{\rm tot}$ be the total
mass within $r_{0}$. Then, $\tilde{\rho}$ is the mass density function written as
\begin{equation}
 \tilde{\rho}(r) = \left\{1+\left(\frac{r}{r_{\rm c}}\right)^2\right\}^{-5/2}.
\label{plummer_density}
\end{equation}
$r_{\rm c}$ gives the clumping scale of the model.
Since we are interested in the motion of S0-2, we take $r_0$ 
as the apoapsis distance of S0-2 ($\sim 0.01\>{\rm pc}$).
In \citet{M05}, they constructed a multi Plummer profile, and the 
most inner core has that $r_{\rm c}=0.015\>{\rm pc}$, 
and therefore we use it for our calculation.

\begin{figure}[t]
  \begin{center}
    \includegraphics[width=80mm]{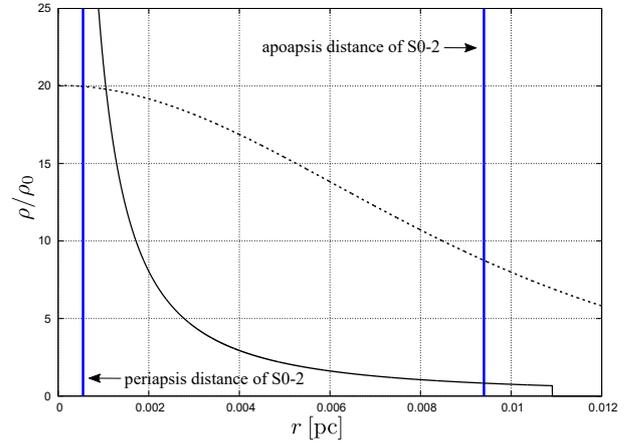}
  \end{center}
  \caption{
    The mass density profiles of the power-law and the Plummer models used in this
    paper. The mass density is normalized by 
    $\rho_0=\eta M_{\rm tot}/(4\pi r_{\rm c}^3)$.
    The solid and the dashed lines represent the mass density profiles of the power-law 
    and the Plummer models. The horizontal axis shows the distance from 
    the Galactic Center. We also show the periapsis, $r_{\rm peri}$, and 
    the apoapsis, $r_{\rm apo}$, distances with the vertical lines. 
    We calculate those with the parameters of S0-2 and Sgr A* 
    in table \ref{Tab.S19} and obtain 
  $r_{\rm peri}=5.5\times10^{-4}\>{\rm pc}$ and 
  $r_{\rm apo}=9.4\times10^{-3}\>{\rm pc}$. (Color online)}
  \label{fig:density}
\end{figure}

\begin{figure}[t]
  \begin{center}
    \includegraphics[width=80mm]{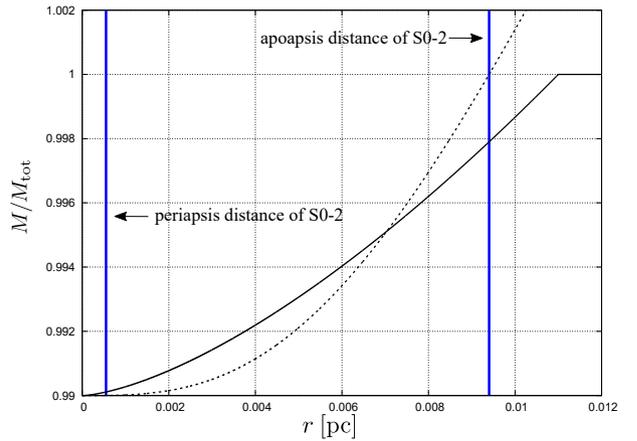}
  \end{center}
  \caption{The mass functions of the power-law and the Plummer model in the 
  case of $\eta=0.01$. The lines are drawn in the same manner 
  in figure \ref{fig:density}. Note that the radius giving $M_{\rm tot}$ is different 
  between the power-law and the Plummer models here. 
  For the power-law model, it is $r_{\rm c}=0.011\>{\rm pc}$ that is the cutoff
  scale.  For the Plummer model, it is $r_0=9.4\times10^{-3}\>{\rm pc}$ that is 
 the apoapsis distance of S0-2. (Color online)}\label{fig:mass}
\end{figure}

Taking $\eta = 0$, both the power-law and the Plummer models revisit
the point mass model with $M_{\rm BH}=M_{\rm tot}$.
It is worth to show the mass density profiles of the power-law and the Plummer models.
The mass density profile for a spherically symmetric mass distribution is given by 
\begin{equation}
  \rho(r) = \frac{1}{4\pi r^2}\frac{dM}{dr}.
\end{equation}
For the power-law model, we have
\begin{equation}
  \rho(r)
  =\frac{\eta M_{\rm tot}}{4\pi r_{\rm c}^3(3-\gamma)}
  \left(\frac{r}{r_{\rm c}}\right)^{-\gamma}. 
\end{equation}
Then, the mass density function of the Plummer model is given by 
\begin{equation}
  \rho(r)
  =\frac{\eta M_{\rm tot}}{4\pi\int_0^{r_{0}}\tilde{\rho}(\xi)\xi^2d\xi}
  \left\{1+\left(\frac{r}{r_{\rm c}}\right)^2\right\}^{-5/2}.
\end{equation}
We show the mass density profiles of the power-law and the Plummer models 
in figure \ref{fig:density}. Moreover, we also show the mass functions in the case
of $\eta=0.01$ in figure \ref{fig:mass}.
\section{Influence on the redshift due to a dark mass distribution}
\label{sec:3}
In this section, we discuss how a dark mass distribution affects the redshift 
of photons from S0-2.
The velocity of S0-2 induced by a dark mass distribution with 1\,\% of the 
mass of Sgr A* is about a tenth of the general relativistic effect \citep{PS09, I11}.
Thus, one thinks that it would be difficult to detect the effect even if we get 
spectroscopic data of S0-2 during the pericenter passage.
However, we should note that the existence of a dark mass distribution 
changes the timing of the pericenter passage because the mass 
affects the period of the star.
Based on Kepler's third law, we can estimate that the change of the period 
due to the dark mass is $\sim 0.01\>{\rm yr}$. 
For S0-2, the time evolution of redshift had varied 
from $4000\>{\rm km\>s^{-1}}$ to 
$-2000\>{\rm km\>s^{-1}}$ within $0.5\>{\rm yr}$ during the pericenter 
passage in 2018.
From that, we can estimate the acceleration of S0-2 in the pericenter passage is 
of order $10^4\>{\rm km\>s^{-1}\>yr^{-1}}$.
Because the timing of the pericenter passage can change $0.01\>{\rm yr}$ 
due to the dark mass, the velocity of S0-2 in the case of the point mass plus an
extended mass model can change, roughly, $\sim 10^2\>{\rm km\>s^{-1}}$ 
from the point mass model at the pericenter.
We will show that the effect on the velocity of S0-2 by the dark mass distribution
reaches $800\>{\rm km\>s^{-1}}$ during the pericenter passage.

\subsection{Redshift in the post-Newtonian approximation}
Let us introduce the redshift of photons from a moving star in the dark mass 
distribution in the context of the post-Newton approximation \citep{D19}.
Taking our coordinate system as a cartesian $(X, Y, Z)$, and assuming
the observer locates on the Z-axis and well far
from the central black hole, the redshift measured at observation time 
$t_{\rm obs}$ is given by
\begin{equation}
  z(t_{\rm obs}) = 
  \frac{v_{\rm Z}(t_{\rm em})}{c}+\frac{v(t_{\rm em})^2}{2c^2}
  +\frac{GM(r(t_{\rm em}))}{c^2 r(t_{\rm em})}
  +\frac{v_{\rm Z0}}{c}.
 \label{redshift}
\end{equation}
$v_{\rm Z}$ is the $Z$-component of the velocity of the star 
${\bf  v}$, and $t_{\rm  em}$ is an emission time of a photon.
The first term represents the radial velocity of the star.
The second and the third terms are the transverse Doppler shift and the 
gravitational redshift, respectively.
The fourth term is the $Z$-component of the solar system's velocity relative to Sgr A*.
For the motion of photons from S0-2 to the observer, Minkowskian treatment is a good 
approximation because general relativistic effects such as the lensing are negligible yet. 
Then, with counting the R$\ddot{\rm o}$mer effect, we have the following relation between
the observation time $t_{\rm obs}$ and the emission time $t_{\rm em}$ with eliminating the 
traveling time from the Galactic Center to the observer:
\begin{equation}
 t_{\rm obs}=t_{\rm em}+\frac{Z(t_{\rm em})}{c},
\end{equation}
where $Z$ is the $Z$-component of the position vector of the star.
We need to search the emission time $t_{\rm em}$ for given $t_{\rm obs}$.
Since $t_{\rm obs} \sim 2010\>{\rm yr}$ and $Z/c \sim 10\>{\rm days}$, 
it is sufficient to use the following relation for our purpose:
\begin{equation}
 t_{\rm em}=t_{\rm obs}-\frac{Z(t_{\rm obs})}{c}.
 \label{time_delay}
\end{equation}
Combing equations (\ref{redshift}) and (\ref{time_delay}), we can
express a model of observed redshift. 

\subsection{Redshift in the point mass model and the point mass plus 
an extended mass model}
\begin{figure}[t]
  \begin{center}
    \includegraphics[width=80mm]{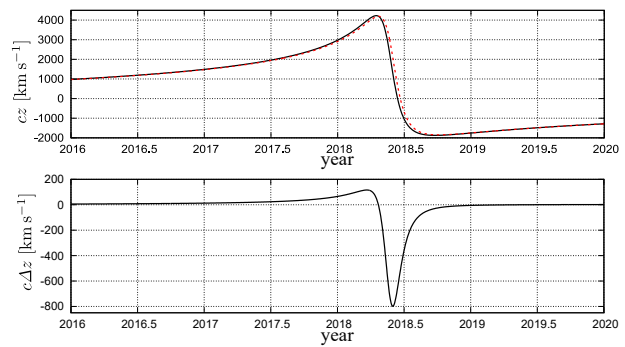}
  \end{center}
  \caption{
    The top panel shows the time evolution of the redshift.
    The solid line represents the case of the point mass model 
    $z_{\rm PM}$ and the dashed line represents the case of 
    the power-law model $z_{\rm PL}$ with $\eta=0.01$.
    The bottom panel shows the discrepancy between the two models 
    that is ${\it \Delta} z=z_{\rm PM}-z_{\rm PL}$. 
    It becomes quite significant during the pericenter passage
    because of the difference in the timing of the pericenter passage.
    The max absolute value reaches $800\>{\rm km\>s^{-1}}$ in the case
    of $\eta=0.01$. 
    (Color online)}\label{fig:redshift}
\end{figure}

We solve equation (\ref{EOM}) numerically for the cases $\eta=0$ (the 
point mass model) and $\eta\neq0$ (the point mass plus an extended mass model).
Because both the power-law model and the Plummer model have a similar 
effect on the time evolution of the redshift, we show the power-law model, for example.
For the point mass model ($\eta=0$), we take the orbital parameters of S0-2 and
the parameters of Sgr A* from the general relativistic best-fitting parameter 
values given in \citet{S19}
(see the row named GR-best-fit in table 5 of the paper).
The general relativistic best-fitting parameter values 
related to calculating the redshift are summarized in table~\ref{Tab.S19}. 
The initial position of S0-2 is set at the previous apocenter at 2010.3383.
On the other hand, for the power-law model $(\eta\neq0)$, 
we take the total mass and the fraction of the extended mass as 
$M_{\rm tot}=4.232\times10^{6}M_{\odot}$ and  $\eta=0.01$, respectively, 
where $M_{\rm tot}$ is equal to the best-fitting parameter value of 
the mass of Sgr A* in the point mass model ($\eta=0$).
The initial conditions are taken as same as the case of the point mass model. 
The results are shown in figure~\ref{fig:redshift}. 
The discrepancy between the two models is quite large during the 
pericenter passage, which reaches $800\>{\rm km\>s^{-1}}$.
We can see that the timing of the maximum of redshift is slightly different
between the cases $\eta=0$ and $\eta\neq0$.
Moreover, the time evolution of the redshift shows the steep 
variation within about $0.5\>{\rm yr}$. 
These cause the large discrepancy during the pericenter passage as shown 
in figure~\ref{fig:redshift}. 

\begin{table}[htbp]
  \caption{Some of the general relativistic best-fitting parameter values 
    of Sgr A* and S0-2 in \citet{S19}.} \label{Tab.S19}
  \begin{tabular}{ccc}\hline
    Parameters ~~&~~ Description~~ &~~ Value \\ \hline
    $M_{\rm BH}$~[$10^6M_{\odot}$]~~&~~Mass of Sgr A*~~&~~4.232 \\
    $R_0$~[{\rm kpc}]~~&~~Distance to Sgr A*~~ &~~8.098 \\
    $v_{Z0}~[{\rm km\>s^{-1}}]$~~&Relative velocity~~&~~-8.345 \\ \hline
    $I$~[{\rm deg}]~~&~~Inclination~~&~~134.239 \\
    $\Omega$~[{\rm deg}]~~&~~Accending node~~&~~227.766 \\
    $\omega$~[{\rm deg}]~~&~~Argument of periapsis~~&~~66.204\\
    $e$~~&~~Eccentricity~~&~~0.8903\\
    $T$~[yr]~~&~~Period~~&~~16.0504\\\hline
  \end{tabular}
\end{table}
\section{Limit on the amount of the extended mass with a simple $\chi^2$ analysis}
\label{sec:4}
We have seen from the figure \ref{fig:redshift} that a dark mass distribution 
affects the time evolution of redshift, and the influence is quite significant during 
the pericenter passage. It can give us a strong constraint on a dark mass 
distribution around Sgr A*, and we will show that in this section.
One of the main points of our paper is that we do not perform the best-fitting 
parameter search to the motion of S0-2 including the dark mass distribution which needs 
much numerical cost as in \citet{Grav18}, \citet{D19}, and \citet{Grav20}.
Instead of the best-fitting parameter search,
we suggest a simple $\chi^2$ analysis to the redshift of S0-2.
The present observational data of S0-2 are well explained by the point mass model.
Thus, the effect  of the dark mass distribution to the motion of S0-2 can be
regarded as a small perturbation within the uncertainties of the observed data.
By applying our $\chi^2$ analysis to the observed redshifts used in \citet{S19}, 
we can easily distinguish between the point mass model and the point mass plus 
an extended mass model thanks to the steep 
variation of the redshift during the pericenter passage. 
As a result, our $\chi^2$ analysis gives
a strong constraint on the amount of the extended mass, which is 
less than $0.5\,\%$ $(\sim 2\times 10^4M_{\odot})$ of Sgr A*.
Our results are comparable to the previous works with the best-fitting parameter
search \citep{Grav18, D19, Grav20}. We will show our $\chi^2$ analysis and 
the results here.

\subsection{Simple $ \chi^2$ analysis}
We calculate the $\chi^2$ for the observed redshifts as
\begin{equation}
  \chi^2= \sum_{i=1}^{N}\frac{(z_i-z_{\rm model}(t_i))^2}{\sigma_i^2},
\end{equation}
where $z_i$ is the observed redshift with the uncertainty
$\sigma_i$, $z_{\rm model}(t_i)$ is a theoretical redshift at the 
observational time $t_i$ and $N$ is the total number of spectroscopic data.
We have the general relativistic best-fitting model of the motion of S0-2 
obtained by \citet{S19}, which corresponds to the case of $\eta=0$.
Moreover, the effect of the extended mass around Sgr A* (the case of $\eta\neq0$)
is allowed within the uncertainties of observational data. 
Therefore, we can distinguish these models by comparing $\chi^2$ 
between the cases of $\eta=0$ and $\eta\neq0$.
Here, let us introduce a normalized $\chi^2$ defined by 
\begin{equation}
  \chi_{\rm n}^2:=\frac{\chi^2_{\eta\neq0}}{\chi^2_{\eta=0}},
\end{equation}
where  $\chi^2_{\eta=0}$ and $\chi^2_{\eta\neq0}$ are the values of $\chi^2$
in the cases of $\eta=0$ and $\eta\neq0$, respectively.
This normalized $\chi^2$ shows the discrepancy between the point mass 
model ($\eta=0$) and the point mass plus 
an extended mass model ($\eta\neq0$).
In our analysis, we use $\chi^2_{\rm n}$ to see whether the model 
with $\eta\neq0$ is acceptable.
To determine the acceptable value of $\chi^2_{\rm n}$,
we calculate $\chi^2_{\rm n}$ with the Newtonian model of S0-2's motion 
and the observed redshifts in \citet{G17} where the spectroscopic data before 
the pericenter passage in 2018 are used.
They have obtained the upper limit of the amount of the extended mass 
inside the orbit of S0-2 is $1\,\%$ of Sgr A*. 
Therefore, by calculating $\chi^2_{\rm n}$ with their upper limit case, 
we can determine the acceptable upper limit of $\chi^2_{\rm n}$ 
based on the spectroscopic data before 2018.
Then, we calculate $\chi^2_{\rm n}$ with the results in \citet{S19} where
the spectroscopic data during the pericenter passage in 2018 are included.
Since the acceptable upper limit of $\chi^2_{\rm n}$ is determined 
based on the observed data before 2018, 
our analysis with the results in \citet{S19} shows how the spectroscopic data 
during the pericenter passage in 2018 affect the constraint on the dark mass distribution.
\subsection{Determination of the acceptable $\chi^2_{\rm n}$
with the observed redshifts before the pericenter passage in 2018}
\begin{figure}[htbp]
  \begin{center}
    \includegraphics[width=80mm]{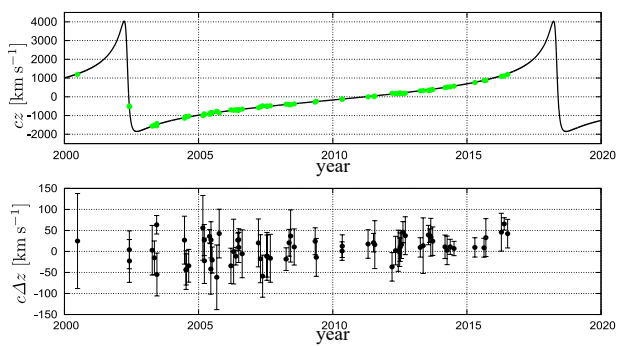}
  \end{center}
  \caption{
    The top panel shows the time evolution of redshift in the 
    case of the point mass model ($\eta=0$). 
    We also show the observed redshifts from 2000--2016 shown in \citet{G17} 
    by the filled circles with error bars.
    The total number of the data, $N$, is 68. The bottom panel is the 
    residual between the observed redshifts and the theoretical model. 
    We can reproduce the time evolution of redshift shown in
    \citet{G17} well. (Color online)}\label{fig:rsG17}
\end{figure}

\begin{figure}[htbp]
  \begin{center}
    \includegraphics[width=70mm]{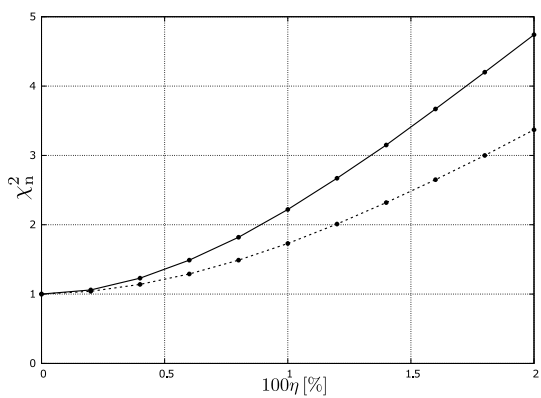}
  \end{center}
  \caption{
    $\chi^2_{\rm n}$ based on the results in \citet{G17}. 
    The solid and dashed lines with points represent $\chi^2_{\rm n}$
    the cases of the power-law and the Plummer models, respectively.
    We calculate $\chi^2_{\rm n}$ at the points and connect them with
    a line for the two models.
    There is no significant difference between them and 
    $1\,\%$ $(\eta=0.01)$ which is the upper limit in \citet{G17}
    corresponds to $\chi^2_{\rm n} \sim 2.0$.}\label{fig:chi2rG17}
\end{figure}

To determine the acceptable range of $\chi^2_{\rm n}$,
we calculate $\chi^2_{\rm n}$ with the results in \citet{G17}.
They have performed an orbital fitting method to the motion of S0-2 
in the context of Newton's gravity both in the point mass model 
and the point mass plus an extended mass model. 
Therefore, we solve equation (\ref{EOM}) eliminating the 2nd and 3rd terms 
in the right-hand side with the parameters in the row 9 in table 1
and the row named S2 in table 3 in \citet{G17}. 
We summarize some of those parameters in table \ref{Tab.G17}. 
We start our calculation from the previous apocenter, which is in 2010.33. 
Figure \ref{fig:rsG17} shows the time evolution of the redshift in the case 
of $\eta=0$ and the residual from the observed redshifts used in \citet{G17}.
We calculate $\chi^2_{\eta=0}$ with those observed redshifts not including 
the observed data during the pericenter passage in 2018.
Then, we also solve the equation of motion in the case of $\eta\neq0$ with 
the initial conditions as same as the case of $\eta=0$, and calculate
$\chi^2_{\eta\neq0}$ for various value of $\eta$. 
Then, we obtain $\chi^2_{\rm n}$.
The results are shown in figure \ref{fig:chi2rG17}. There is no
significant difference between the power-law (the solid line 
with points in figure \ref{fig:chi2rG17}) and the Plummer (the dotted
line with points in figure \ref{fig:chi2rG17}) models.
In \citet{G17}, they concluded that the amount of the 
extended mass is $1\,\%$ of Sgr A*, i.e., $\eta=0.01$ at most.
This corresponds to $\chi^2_{\rm n}$ of $1.7$ to $2.2$ in figure \ref{fig:chi2rG17}.
Therefore, we can give the upper value of the acceptable range of $\chi^2_{\rm n}$ as 
$\chi^2_{\rm n}\sim 2.0$.
\begin{table}[htbp]
  \caption{Some of the best-fitting parameter values of 
    Sgr A* and S0-2 in \citet{G17}.} \label{Tab.G17}
  \begin{tabular}{ccc}\hline
    Parameters ~~&~~ Description~~ &~~ Value \\ \hline
    $M_{\rm BH}$~[$10^6M_{\odot}$]~~&~~Mass of Sgr A*~~&~~4.28 \\
    $R_0$~[{\rm kpc}]~~&~~Distance to Sgr A*~~ &~~8.32 \\
    $v_{Z0}~[{\rm km\>s^{-1}}]$~~&Radial velocity of Sgr A*~~&~~14.2
    \\ 
    \hline
    $I$~[{\rm deg}]~~&~~Inclination~~&~~134.18 \\
    $\Omega$~[{\rm deg}]~~&~~Accending node~~&~~226.94 \\
    $\omega$~[{\rm deg}]~~&~~Argument of periapsis~~&~~65.51\\
    $e$~~&~~Eccentricity~~&~~0.8839\\
    $T$~[yr]~~&~~Period~~&~~16.00\\ \hline
  \end{tabular}
\end{table}


\subsection{Limit on the amount of the extended mass including Subaru observational data during the pericenter passage in 2018}
\begin{figure}[htbp]
  \begin{center}
    \includegraphics[width=70mm]{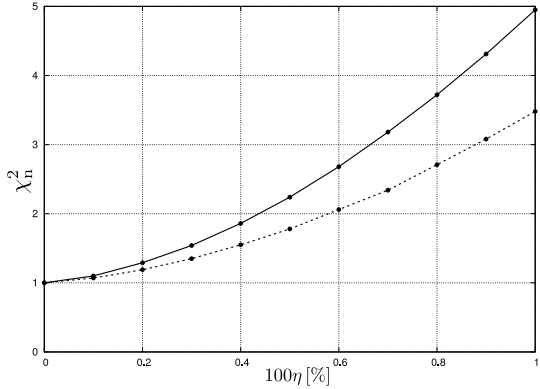}
  \end{center}
  \caption{ $\chi^2_{\rm n}$ based on the results in \citet{S19}. 
    The solid and dashed lines with points represent $\chi^2_{\rm n}$ 
    in the cases of the power-law and the Plummer models, respectively.
    We calculate $\chi^2_{\rm n}$ at the points and connect them with
    a line for the two models.
    Focusing the case of $100\eta=1\,\%$ which is acceptable in 
    \citet{G17}, $\chi^2_{\rm n}>3$ for the both models.
    Since the upper value of $\chi^2_{\rm n}$ obtained in section 4.2 
    is $\chi^2_{\rm n}\sim2.0$, we exclude the case of $100\eta=1\,\%$.
    It means that the acceptable amount of extended mass is bounded 
    less than $1\,\%$ of Sgr A* due to adding the data from Subaru/IRCS.
  }\label{fig:chi2Saida}
\end{figure}

Let us calculate $\chi^2_{\rm n}$ with the results in \citet{S19}.
They have found the general relativistic best-fitting orbital model 
of S0-2 using the astrometric and the spectroscopic data given in 
\citet{B16} and \citet{G17} and the spectroscopic data obtained by Subaru/IRCS.
We expect that the spectroscopic data during the pericenter passage in 2018 obtained by
Subaru/IRCS would bound the amount of the extended mass more strongly
than the bound given in \citet{G17}.
We numerically solve equation (\ref{EOM}) with the parameters in 
table~{\ref{Tab.S19}} and calculate $\chi_{\rm n}^2$ for various $\eta$. 
The results are shown in figure \ref{fig:chi2Saida}.
Obeying our criterion obtained in section 4.2 that is $\chi^{2}_{\rm n}\leq2.0$, 
the acceptable fraction of the mass is about 
$100\eta\sim0.5\,\%\sim2\times10^4M_{\odot}$ in the both models,
which is a stronger constraint than that given in \citet{G17}.

\begin{figure}[htbp]
  \begin{center}
    \includegraphics[width=70mm]{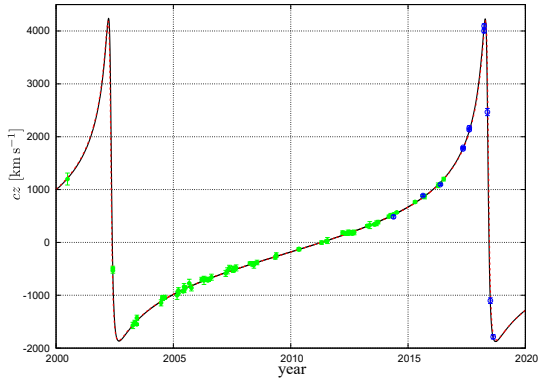}
  \end{center}
  \caption{
    The time evolution of redshift.
    The solid and dashed line represent the cases $\eta=0$ and
    $\eta=0.01$ in the power-law model, respectively.
    The points with error bars are the observed redshifts. 
    The filled circles (green) are from \citet{G17}.
    The open circles (blue) are the data obtained by Subaru/IRCS \citep{S19}. 
    The total number of data, $N$, is 80.  (Color online)} \label{fig:rs00to20}
\end{figure}

\begin{figure}[htbp]
  \begin{center}
    \includegraphics[width=70mm]{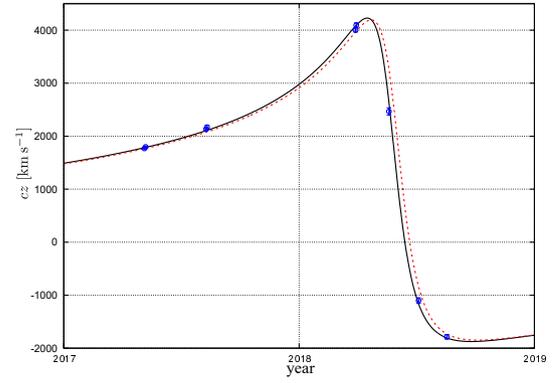}
  \end{center}
  \caption{
    The time evolution of redshift around the pericenter passage. 
    We can see that the data points are well fitted in the case of 
    $\eta=0$ but the case of $\eta=0.01$ is off the points during 
    the pericenter passage. (Color online)}\label{fig:rs17to19}
\end{figure}

\begin{figure}[htbp]
  \begin{center}
    \includegraphics[width=70mm]{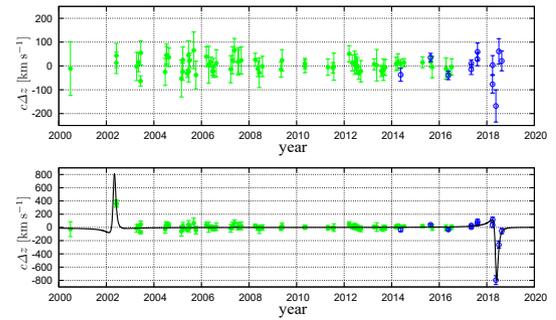}
  \end{center}
  \caption{
    The residual between the observed redshifts used in \citet{S19} and
    the theoretical models.
    The top and the bottom panels show the cases of $\eta=0$ and 
    $\eta=0.01$ in the power-law model, respectively.
    In the bottom panel, we plot $c(z_{\rm PM}-z_{\rm PL})$ in
    figure \ref{fig:redshift} again with the solid line.  
    We can see that the spectroscopic data with Subaru/IRCS during the 
    pericenter passage in 2018 play an important role to constrain 
    the amount of the extended mass.(Color online)}\label{fig:dzSaida}
\end{figure}

To better understand our results, let us see the detail of the cases
of $\eta=0$ and $\eta=0.01$.
Since there is no significant difference between the power-law
and the Plummer models, we only show the power-law model.
Figure \ref{fig:rs00to20} shows the time evolution of redshift in the
cases of $\eta=0$ and $\eta=0.01$ in the power-law model.
The solid and the dashed lines represent the cases $\eta=0$ and 
$\eta=0.01$, respectively.
The points with error bars represent the observed redshifts used in \citet{S19}.
The filled circles (green) are the observed redshifts in \citet{G17}, and
the open circles (blue) are those from Subaru/IRCS shown in \citet{S19}.
We zoom that figure around the pericenter passage in 2018 and display it in
figure \ref{fig:rs17to19}.
Moreover, we show the residual between the observed redshifts and the 
theoretical models in figure \ref{fig:dzSaida}.

Our analysis shows that the redshift observations around the pericenter 
passage are crucial to constrain the amount of the extended mass around Sgr A*. 
As shown in figure \ref{fig:dzSaida}, the redshift difference between 
$\eta=0$ and $\eta=0.01$ is very small between 2003 and 2017. 
On the other hand, the difference has become as large as $800\>{\rm km\>s^{-1}}$ 
in 2018, during the pericenter passage of S0-2.  
This significant difference was thus expected to be detected in the 2018 observations. 
By comparing the top and bottom panels in figure \ref{fig:dzSaida}, 
we conclude that the extended mass with $1\,\%$ of Sgr A* is excluded. 
The observed redshifts and the point mass model $(\eta=0)$ are consistent 
within $2\sigma$ (the top panel in figure \ref{fig:dzSaida}). 
By contrast, the residuals between the observations and the point mass plus 
the extended mass model $(\eta=0.01)$ are more 
than several $100\>{\rm km\>s^{-1}}$ (bottom panel). 
It means that $\eta=0.01$ model can not well reproduce the observed redshifts.
One of the reasons we can clearly differentiate the two models ($\eta=0$ and $\eta=0.01$)
is that we could measure the redshift of S0-2 at 2018.38, the third data point 
from the right in figure \ref{fig:rs00to20}--\ref{fig:dzSaida}. 
This is not the peak or bottom of the redshift, but it is close to 
the maximum point in the difference between the two models 
(the bottom panel in figure \ref{fig:redshift}). 
It will be crucial to make observations at an appropriate time to 
give a stronger constraint on the amount of the extended mass and 
the mass density profile.

\section{Summary and Discussion}
The star S0-2 moving around Sgr A* passed through the pericenter in
May 2018. We expect that this event gives us new information about the
environment around Sgr A*.
In this paper, we have discussed a dark mass distribution surrounding Sgr A* 
inside the orbit of S0-2 (within $\sim 0.01\>{\rm pc}$).
The redshift of photons from S0-2 had varied 
from $4000\>{\rm km\>s^{-1}}$ to $-2000\>{\rm km\>s^{-1}}$ 
during the pericenter passage in 2018, which is within $0.5\>{\rm yr}$.
This steep variation gives a strong constraint on the dark mass distribution.
We suggested a simple $\chi^2$ analysis for the redshift of S0-2
to constrain the dark mass distribution and applied to the results in \citet{S19}.
As a result, thanks to the steep variation of the redshift during the 
pericenter passage in 2018, we can bound the amount of the extended mass inside 
the orbit of S0-2 less than $0.5\,\%$ ($\sim 2\times10^4M_{\odot}$) of 
the mass of Sgr A*. Our constraint is stronger by the factor $1/2$ than
the result in \citet{G17} where the observational data in 2018 were 
not included. It means that the spectroscopic data during the pericenter 
passage in 2018 is crucial to give that strong constraint on the dark mass
distribution. Furthermore, our results are comparable to the results with
the best-fitting parameter search to the motion of S0-2 including the 
data during the pericenter passage in 2018 \citep{Grav18,D19,Grav20}. 
By focusing on the steep variation in the time evolution of the redshift of S0-2 
during the pericenter passage, we can give a constraint on the amount 
of the extended mass with less numerical costs.
Although we consider both the power-law and the Plummer models 
as for the dark mass distribution, we do not find the significant difference
between them. 

Both the power-law and the Plummer models represent a stellar cluster 
surrounding Sgr A*.
A dark matter distribution surrounding Sgr A* is also available and considered
\citep{GS99, Sad13}. 
The motion of S-stars can bound a dark matter profile.
The amount of the dark matter within the orbit of S0-2 
is a few \% of the mass of Sgr A* at most \citep{HG06, L18}. 
The observational data of S0-2 in 2018 were not included in these works.
Because there is no significant difference between the mass profile 
functions of a stellar cluster and a dark matter models, 
the upper limit of the amount of the extended mass
would be less than $1\,\%$ of Sgr A* even if one considers 
a dark matter component. 

In this paper, we have focused on the spectroscopic data of S0-2. 
The other observable is the astrometry of S0-2, and it gives us 
information about the dark mass distribution, of course.
\citet{R01} pointed out that the existence of dark mass 
distribution raises the apocenter shift even if in Newton's gravity. 
That is comparable to the relativistic apocenter shift 
if the amount  of the extended mass is less than $1\,\%$ 
of the mass of Sgr A*.
\citet{Z07} studied how a dark matter profile affects 
the apocenter shift for S0-2 and showed that the astrometric 
observations could exclude some dark matter profiles.
\citet{Grav19} considered a scalar field cloud surrounding Sgr A* 
as a dark matter model and studied the motion of S0-2.
They showed that the motion of S0-2 is sensitive
to the width of a dark matter cloud. 
It gives the range of the mass of the dark matter particle.
Moreover, we can consider an exotic system 
that is without the central black hole, e.g., a dark matter core 
\citep{BM19}, a naked singularity \citep{Dey19}.
We can not exclude these alternatives in the present observations of S0-2. 
The $\chi^2$ analysis suggested in this paper would be useful to 
give a constraint on these alternatives.

Recently, \citet{Grav20} reported that they had detected the pericenter shift for S0-2, 
which is consistent with the general relativity.
S0-2 will experience the apocenter passage in 2026, and we expect to get 
new information through the event.
Furthermore, \citet{P20} has found a new S-star moving 
inside the orbit of S0-2, whose period is about $10\,{\rm yr}$. 
These new observational data would be useful to constrain on 
the dark mass distribution or to resolve the components of the dark mass 
distribution surrounding Sgr A* and to test alternatives for a black hole.
For a research of a dark mass distribution surrounding Sgr A*, 
we should prepare a general relativistic dark mass model including 
the higher-order terms than the first post-Newtonian term.
For example, if a geodesic represents the orbit of a star even 
for the case $M=M(r)$, we expect that new terms proportional to 
$\partial_r M$ appear in equation (\ref{EOM}), which we do not 
introduce in this paper.
$\partial_r M$  could be comparable with $(M-M_{\rm tot})/r$, that is, 
the mass profile correction due to the dark mass. 
Therefore, it is worth including these new relativistic terms to evaluate 
the evolution of the star more precisely. 
We will investigate this issue in the near future.
\bigskip
\begin{ack}
Y.~T. was supported by JSPS KAKENHI, Grant-in-Aid for Young Scientists (B)
26800150.
S.~N. was supported by JSPS KAKENHI, 
Grant-in-Aid for Challenging Exploratory Research18K18760, and 
Grant-in-Aid for Scientific Research (A) 19H00695. 
T.~O. was supported by JSPS KAKENHI, Grantin-
Aid for JSPS fellows JP17J00547.
H.~S. was supported by JSPS KAKENHI, Grant-in-Aid for Challenging 
Exploratory Research 26610050, and Grant-in-Aid for Scientific Research 
(B) 19H01900. 
M.~T. was supported by DAIKO FOUNDATION, and JSPS KAKENHI,
Grant-in-Aid for Scientific Research (C) 17K05439.
\end{ack}


\end{document}